\documentclass[epj]{svjour}
\usepackage{graphics}
\begin{document}
\title{Phase diagram of the Kohn-Luttinger superconducting state for bilayer graphene}
\author{Maxim Yu. Kagan\inst{1,2}\thanks{\emph e-mail: kagan@kapitza.ras.ru} \and Vitaly A. Mitskan\inst{3,4} \and Maxim M. Korovushkin\inst{3}% etc
}                     % Do not remove
\institute{P.\,L. Kapitza Institute for Physical Problems, Russian
Academy of Sciences, Moscow 119334, Russia \and National Research
University Higher School of Economics, Moscow 109028, Russia \and
L.\,V. Kirensky Institute of Physics, Siberian Branch of Russian
Academy of Sciences, 660036 Krasnoyarsk, Russia \and M.\,F.
Reshetnev Siberian State Aerospace University, 660014 Krasnoyarsk,
Russia }
\date{Received: date / Revised version: date}
\abstract{The effect of the intersite and interplane Coulomb
interactions between the Dirac fermions on the formation of the
Kohn-Luttinger superconductivity in bilayer doped graphene is
studied disregarding the effects of the van der Waals potential of
the substrate and both magnetic and non-magnetic impurities. The
phase diagram determining the boundaries of superconductive
domains with different types of symmetry of the order parameter is
built using the extended Hubbard model in the Born weak-coupling
approximation with allowance for the intratomic, interatomic, and
interlayer Coulomb interactions between electrons. It is shown
that the Kohn-Luttinger polarization contributions up to the
second order of perturbation theory in the Coulomb interaction
inclusively and an account for the long-range intraplane Coulomb
interactions significantly affect the competition between the
superconducting $f-$, $p+ip-$, and $d+id-$wave pairings. It is
demonstrated that the account for the interplane Coulomb
interaction enhances the critical temperature of the transition to
the superconducting phase.
 % end of PACS codes
} %end of abstract
%

%
%\PACS{
%      {74.20.Mn}{Superconductivity: nonconventional mechanisms} \and
%      {74.25.Dw}{Superconductivity: phase diagrams} \and
%      {81.05.ue}{Carbon-based materials: graphene}
%     } % end of PACS codes
%} %end of abstract
%%
\authorrunning{Kagan, Mitskan, Korovushkin}
\maketitle

\section{Introduction}
\label{intro} In recent years, there has been an increased
interest in the possibility of the development of the Cooper
instability in graphene under appropriate experimental conditions.
Although so far this possibility has not been confirmed, it was
experimentally
shown~\cite{Heersche07,Shailos07,Du08,Ojeda09,Kanda10,Han14} that
graphene becomes superconducting when it is in a contact with
ordinary superconductors. This fact stimulated theoretical studies
on possible implementation of the superconducting phase in an
idealized monolayer and bilayer graphene where the authors did not
take into account the effect of nonmagnetic impurities and van der
Waals potential of the substrate.

Along with the numerous studies of this problem using the
electron-phonon
mechanism~\cite{Kopnin08,Basko08,Lozovik10,Einenkel11,Classen14},
pairing mechanisms caused by electron
correlations~\cite{Black07,Honerkamp08,Vucicevic12,Milovanovic12},
and other exotic superconductivity
mechanisms~\cite{Hosseini12a,Hosseini12b}, some authors widely
discuss the possibility of the development of Cooper instability
in the above-mentioned systems using the Kohn-Luttinger
mechanism~\cite{Kohn65}, which suggests the emergence of
superconducting pairing in the systems with the purely repulsive
interaction~\cite{Fay68,Kagan88,Kagan89,Baranov92,Kagan14a}.

As it was shown in~\cite{Gonzalez08}, the Cooper instability can
occur in an idealized graphene single layer due to the strong
anisotropy of the Fermi contour for Van Hove filling $n_{VH}$,
which, in fact, originates from the Kohn-Luttinger mechanism.
According to the results obtained in~\cite{Gonzalez08}, this
Cooper instability in graphene evolves predominantly in the
$d-$wave channel and can be responsible for the critical
superconducting transition temperatures up to $T_c\sim10\,K$,
depending on the proximity of the chemical potential level to the
Van Hove singularity. The theoretical analysis of the competition
between the ferromagnetic and superconducting instabilities
showed~\cite{McChesney10} that the tendency to superconductivity
due to strong modulation of the effective interaction along the
Fermi contour, i.e., due to electron-electron interactions alone,
prevails. In this case, the superconducting instability evolves
predominantly in the $f$-wave channel.

The competition between the Kohn-Luttinger superconducting phase
and the spin density wave phase at the Van Hove filling and near
it in the graphene single layer was analyzed
in~\cite{Nandkishore12,Kiesel12} using the functional
renormalization group method. It was found that superconductivity
with the $d+id-$wave symmetry of the order parameter prevails in a
large domain near the Van Hove singularity, and a change in the
calculated parameters may lead to a transition to the phase of the
spin density wave. According to~\cite{Kiesel12}, far away from the
Van Hove singularity, the long-range Coulomb interactions change
the form of the $d+id-$wave function of a Cooper pair and can
facilitate superconductivity with the $f-$wave symmetry of the
order parameter. The competition between the superconducting
phases with different symmetry types in the wide electron density
range $1<n\leq n_{VH}$ in the graphene single layer was studied
in~\cite{Kagan14,Nandkishore14}. It was demonstrated that at
intermediate electron densities the long-range Coulomb
interactions facilitate implementation of superconductivity with
the $f-$wave symmetry of the order parameter, while at approaching
the Van Hove singularity, the superconducting pairing with the
$d+id-$symmetry type evolves~\cite{Kagan14,Nandkishore14}.

The conditions for the Kohn-Luttinger superconducting pairing was
analyzed also in graphene bilayer
\cite{Vafek10a,Vafek10b,Guinea12,Vafek14}. According to the
results of~\cite{Gonzalez13}, the ferromagnetic instability near
the Van Hove singularities dominates over the Kohn-Luttinger
pairing in graphene bilayer. It should be noted, however, that in
these calculations only the Coulomb repulsion of electrons on one
site was taken into account. Authors of~\cite{Hwang08} calculated
the screening function of Coulomb interaction in the doped and
undoped bilayer graphene in the random phase approximation (RPA).
They established that the static polarization operator in the
doped regime contains the singular part (the Kohn anomaly) that
significantly exceeds one calculated for monolayer or 2D electron
gas. As it is known, the Kohn anomaly~\cite{Migdal58,Kohn59}
facilitates the effective attraction between two particles,
inducing a contribution that always exceeds the repulsive
contribution connected with the regular part of the polarization
operator for the angular momenta $l\neq0$ of two
particles~\cite{Kohn65}. Therefore, one can expect that the
critical superconducting temperature $T_c$ in an idealized bilayer
can exceed the corresponding value for graphene monolayer.

Additionally, it was shown in papers~\cite{Kagan91,KaganValkov11a}
that the value of $T_c$ can be increased in the framework of the
Kohn-Luttinger mechanism even for low carrier densities if the
spin-polarized two-band situation or a multilayer system is
considered. In this situation, the role of the pairing spins "up"
is played by electrons of one band (layer), while the role of the
screening spins "down" is played by electrons of another band
(layer). Coupling between the electrons from the two bands occurs
owing to the interband (interlayer) Coulomb interaction. In this
case, the following mechanism is possible: electrons of one sort
form a Cooper pair by polarizing the electrons of another
sort~\cite{Kagan91,KaganValkov11a}. This mechanism can be realized
also in quasi-2D systems.

\begin{figure}
\begin{center}
\resizebox{0.48\textwidth}{!}{%
  \includegraphics{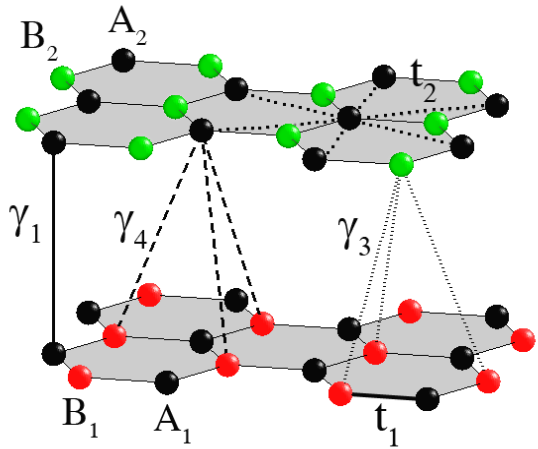}
} \caption{Crystalline structure of bilayer graphene. Atoms $A_1$
and $B_1$ in the lower monolayer are shown by red and black
circles; black and green circles in the upper layer correspond to
atoms $A_2$ and $B_2$. Intraplane electron hoppings are labeled by
$t_1$ and $t_2$; $\gamma_1,\,\gamma_3$ and $\gamma_4$ show the
interlayer hoppings.}\label{bilayer_structure}
\end{center}
\end{figure}

In this paper, in the Born weak-coupling approximation, we
consider the Kohn-Luttinger superconducting pairing in an
idealized graphene bilayer. We calculate the phase diagram, which
reflects the competition between the superconducting phases with
different types of the symmetry of the order parameter, taking
into account the second-order contributions in the Coulomb
interaction to the effective interaction of electrons in the
Cooper channel. We analyze modification of the phase diagram with
allowance for the Coulomb repulsion between electrons of the same,
of the nearest, and of the next-to-nearest carbon atoms in a
single layer, as well as the interlayer Coulomb interactions. We
demonstrate the importance of taking into account the Coulomb
repulsion of electrons on different crystal lattice sites and in
different layers of bilayer graphene. The account of Coulomb
repulsion changes the phase diagram of the superconducting state
and, under certain conditions, increases the critical temperature.

\section{Theoretical model}

We consider an idealized graphene bilayer, assuming that two
layers are arranged in accordance with the $AB$ type, i.e., one
layer is rotated on 60$^o$ relative to the other
one~\cite{McCann06,McCann13}. Let us choose the arrangement of the
sublattices in the layers in such a way that the sites from
different layers located one above another belong to the
sublattices $A_1$ and $A_2$ respectively, while the other sites
belong to the sublattices $B_1$ and $B_2$
(Fig.~\ref{bilayer_structure}). In the Shubin-Vonsovsky (extended
Hubbard) model~\cite{Shubin34}, the Hamiltonian for the graphene
bilayer which takes into account electron hoppings between the
nearest and next-to-nearest atoms, as well as the Coulomb
repulsion between electrons of the same and of the adjacent atoms
and the interlayer Coulomb interaction of electrons, in the
Wannier representation has the form:
\begin{eqnarray}\label{HamiltonianBilayer}
\hat{H}&=&\hat{H}_0+\hat{H}_{int},\\
\hat{H}_0&=&(\varepsilon-\mu)\Biggl(\sum_{if\sigma}\hat{n}^{A}_{if\sigma}+
\sum_{ig\sigma}\hat{n}^{B}_{ig\sigma}\Biggr)\nonumber\\
&-&t_1\sum_{f\delta\sigma}(a^{\dag}_{1f\sigma}b_{1,f+\delta,\sigma}+
a^{\dag}_{2f\sigma}b_{2,f-\delta,\sigma}+\textrm{h.c.})\nonumber\\
&-&t_2\sum_{i\sigma}\Biggl(\sum_{\langle\langle
fm\rangle\rangle}a^{\dag}_{if\sigma}a_{im\sigma}+\sum_{\langle\langle
gn\rangle\rangle}b^{\dag}_{ig\sigma}b_{in\sigma}+
\textrm{h.c.}\Biggr)\nonumber\\
&-&\gamma_1\sum_{f\sigma}(a^{\dag}_{1f\sigma}a_{2f\sigma}+\textrm{h.c.})\nonumber\\
&-&\gamma_3\sum_{g\delta\sigma}(b^{\dag}_{1g\sigma}b_{2,g+\delta,\sigma}+\textrm{h.c.})\nonumber\\
&-&\gamma_4\sum_{f\delta\sigma}(a^{\dag}_{1f\sigma}b_{2,f-\delta,\sigma}+
a^{\dag}_{2f\sigma}b_{1,f+\delta,\sigma}+\textrm{h.c.}),\label{H0Bilayer}\\
\hat{H}_{int}&=&U\biggl(\sum_{if}
\hat{n}^{A}_{if\uparrow}\hat{n}^{A}_{if\downarrow}+ \sum_{ig}
\hat{n}^{B}_{ig\uparrow}\hat{n}^{B}_{ig\downarrow}\biggr)+\nonumber\\
&+&V_1\sum_{f\delta\sigma\sigma'}
\Bigl(\hat{n}^{A}_{1f\sigma}\hat{n}^{B}_{1,f+\delta,\sigma'}+
\hat{n}^{A}_{2f\sigma}\hat{n}^{B}_{2,f-\delta,\sigma'}\Bigr)+\nonumber\\
&+&\frac{V_2}{2}\sum_{i\sigma\sigma'}\Biggl(\sum_{\langle\langle
fm\rangle\rangle}\hat{n}^{A}_{if\sigma}\hat{n}^{A}_{im\sigma'}+\sum_{\langle\langle
gn\rangle\rangle}\hat{n}^{B}_{ig\sigma}\hat{n}^{B}_{in\sigma'}\Biggr)+\nonumber\\
&+&G_1\sum_{f\sigma\sigma'}
\hat{n}^{A}_{1f\sigma}\hat{n}^{A}_{2f\sigma'} +
G_3\sum_{g\delta\sigma\sigma'}
\hat{n}^{B}_{1g\sigma}\hat{n}^{B}_{2,g+\delta,\sigma'}+\nonumber\\
&+&G_4\sum_{f\delta\sigma\sigma'}
\Bigl(\hat{n}^{A}_{1f\sigma}\hat{n}^{B}_{2,f-\delta,\sigma'}+
\hat{n}^{A}_{2f\sigma}\hat{n}^{B}_{1,f+\delta,\sigma'}\Bigr).\label{HintBilayer}
\end{eqnarray}
In (\ref{HamiltonianBilayer})--(\ref{HintBilayer}), the operators
$a^{\dag}_{1f\sigma}(a_{1f\sigma})$ create (annihilate) an
electron with the spin projection $\sigma=\pm1/2$ at site $f$ of
the sublattice $A_1$; $\hat{n}^{A}_{1f\sigma}=
a^{\dag}_{1f\sigma}a_{1f\sigma}$ denotes the operators of the
numbers of fermions at the $f$ site of the sublattice $A_1$
(analogous notations are used for the sublattices $A_2$, $B_1$,
and $B_2$). Vector $\delta (-\delta)$ connects the nearest atoms
of the hexagonal lattice of the lower (upper) layer. Index $i=1,2$
in Hamiltonian (\ref{HamiltonianBilayer}) denotes the number of
layer. We assume that the one-site energies are identical
($\varepsilon_{Ai}=\varepsilon_{Bi}=\varepsilon$) and the position
of the chemical potential $\mu$ and number of carriers $n$ in
graphene bilayer can be controlled by a gate electric field. In
the Hamiltonian, $t_1$ is the hopping integral between the
neighboring atoms (hoppings between different sublattices), $t_2$
is the hopping integral between the next-to-nearest neighboring
atoms (hoppings in the same sublattice), $U$ is the parameter of
Coulomb repulsion between electrons of the same atom with the
opposite spin projections (Hubbard repulsion), and $V_1$ and $V_2$
are the Coulomb interactions between electrons of the nearest and
the next-to-nearest carbon atoms in a single layer. The symbol
$\langle\langle~\rangle\rangle$ indicates that summation is made
only over next-to-nearest neighbors; the symbols
$\gamma_1,\,\gamma_3,\,\gamma_4$ denote the parameters of the
interlayer electron hoppings (Fig.~\ref{bilayer_structure}), and
$G_1$, $G_3$ and $G_4$ are the interlayer Coulomb interactions
between electrons.

We diagonalize the Hamiltonian $\hat{H}_0$ using the Bogolyubov
transformation
\begin{eqnarray}\label{uv2}
\alpha_{i\vec{k}\sigma}&=& w_{i1}(\vec{k}){a_{1 \vec{k}\sigma }} +
w_{i2}(\vec{k}){a_{2\vec{k}\sigma
}}\\
&+&w_{i3}(\vec{k}){b_{1\vec{k}\sigma }} +
w_{i4}(\vec{k}){b_{2\vec{k}\sigma }},\quad i=1,2,3,4.\nonumber
\end{eqnarray}
As a result, $\hat{H}_0$ acquires the form
\begin{eqnarray}
\hat H_0 =\sum\limits_{i=1}^4 \sum\limits_{ \vec{k}\sigma }
E_{i\vec{k}}
{\alpha_{i\vec{k}\sigma}^{\dag}\alpha_{i\vec{k}\sigma}}.
\end{eqnarray}
Since the results of ab initio calculations for
graphite~\cite{Dresselhaus02,Brandt88} showed a very small value
of the interlayer hopping parameter $\gamma_4$, hereinafter we
assume that $\gamma_4=0$. Then, the four-band energy spectrum of
the graphene bilayer is described by the expressions
\begin{eqnarray}\label{spectra}
&&E_{i\vec{k}}=\varepsilon\pm\sqrt{A_{\vec{k}}\pm\sqrt{B_{\vec{k}}}}-t_2f_{\vec{k}},\\
&&A_{\vec{k}}=\frac14\Bigl(2a^2+4|b_{\vec{k}}|^2+2|d_{\vec{k}}|^2\Bigr),\nonumber\\
&&B_{\vec{k}}=\frac14\Bigl(|d_{\vec{k}}|^2(|d_{\vec{k}}|^2-2a^2+4|b_{\vec{k}}|^2)+a^4+4a^2|b_{\vec{k}}|^2\nonumber\\
&&\qquad+4ab^2_{\vec{k}}d_{\vec{k}}+4ab_{\vec{k}}^{*2}d^*_{\vec{k}}\Bigr),\nonumber\\
&&a=\gamma_1,\quad b_{\vec{k}}=t_1u_{\vec{k}},\quad
d_{\vec{k}}=\gamma_3u_{\vec{k}}\nonumber,
\end{eqnarray}
where the following notation has been introduced:
\begin{eqnarray}\label{f_k}
&&f_{\vec{k}}=2\cos(\sqrt{3}k_y)+
4\cos\biggl(\frac{\sqrt{3}}{2}k_y\biggr)\cos\biggl(\frac{3}{2}k_x\biggr),\\
&&u_{\vec{k}}=\displaystyle\sum_{\delta}e^{i
\vec{k}\delta}=e^{-ik_x}+
2e^{\frac{i}{2}k_x}\cos\biggl(\frac{\sqrt{3}}{2}k_y\biggr),\label{u_k}\\
&&|u_{\vec{k}}|=\sqrt{3+f_{\vec{k}}}.
\end{eqnarray}

\begin{figure}
\begin{center}
\resizebox{0.41\textwidth}{!}{%
  \includegraphics{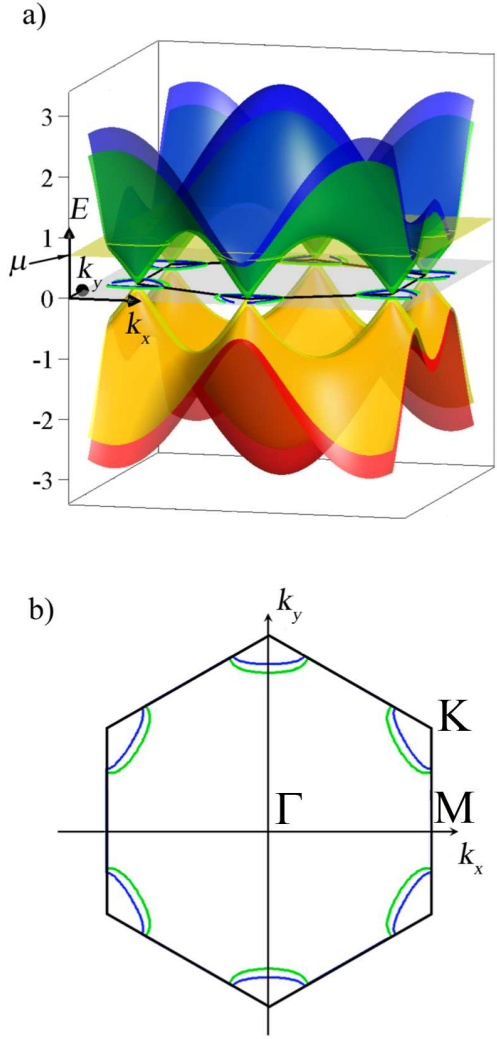}
} \caption{(a) Energy bands of graphene bilayer in the vicinity of
the Dirac points and (b) formation of the Fermi contour at
$t_2=0,\,\gamma_1=0.12,\,\gamma_3=0.1$, and $\mu=0.7$ (all the
parameters are given in units of $|t_1|$).}\label{two_contours}
\end{center}
\end{figure}

In this paper, the conditions for the implementation of the
Kohn-Luttinger superconductivity are analyzed by considering the
situation when upon doping of the graphene bilayer the chemical
potential falls into the two upper energy bands $E_{1\vec{k}}$ and
$E_{2\vec{k}}$ (Fig.~\ref{two_contours}a). Then, if
$\gamma_1\neq0$ and the inequality $\mu>\gamma_1$ is valid, the
Fermi contour will consist of two lines (Fig.~\ref{two_contours}b)
in the vicinity of each Dirac point for the electron densities
$1<n<n_{VH}$, where $n$ is the electron density calculated per
atoms of one layer.

The coefficients of the Bogolyubov transformation can be found
from the system of homogeneous equations
\begin{eqnarray}\label{wz}
\left(%
\begin{array}{cccc}
  x_i & a & b^*_{\vec{k}} & 0 \\
  a & x_i & 0 & b_{\vec{k}} \\
  b_{\vec{k}} & 0 & x_i & d^*_{\vec{k}} \\
  0 & b^*_{\vec{k}} & d_{\vec{k}} & x_i \\
\end{array}%
\right)\left(%
\begin{array}{c}
  w_{i1} \\
  w_{i2} \\
  w_{i3} \\
  w_{i4} \\
\end{array}%
\right)=0,
\end{eqnarray}
where $x_i=E_{i\vec{k}}-\varepsilon+t_2f_{\vec{k}}$.

In the Bogolyubov representation, the Hamiltonian $\hat H_{int}$
(\ref{HintBilayer}) in terms of the operators
$\alpha_{1\vec{k}\sigma},\,\alpha_{2\vec{k}\sigma},\,\alpha_{3\vec{k}\sigma}$
and $\alpha_{4\vec{k}\sigma}$ reads as follows:
\begin{eqnarray}\label{Hint_ab}
\hat H_{int} &=& \frac{1}{N}\sum\limits_{ijlm\sigma\atop
\vec{k}\vec{p}\vec{q}\vec{s}}
\Gamma_{ij;lm}^{||}(\vec{k},\vec{p}|\vec{q},\vec{s})
\alpha_{i\vec{k}\sigma}^\dag \alpha_{j\vec{p}\sigma}^\dag
\alpha_{l\vec{q}\sigma}\alpha_{m\vec{s}\sigma}
\nonumber\\
&\times&\delta (\vec{k}+\vec{p}-\vec{q}-\vec{s})\\
&+& \frac{1}{N}\sum\limits_{ijlm\atop
\vec{k}\vec{p}\vec{q}\vec{s}}\Gamma_{ij;lm}^{\bot}(\vec{k},\vec{p}|\vec{q},\vec{s})
\alpha_{i\vec{k}\uparrow}^\dag \alpha_{j\vec{p}\downarrow}^\dag
\alpha_{l\vec{q}\downarrow}
\alpha_{m\vec{s}\uparrow}\nonumber\\
&\times&\delta (\vec{k}+\vec{p}-\vec{q}-\vec{s}),\nonumber
\end{eqnarray}
where $\delta(x)$ is the Dirac delta-function and
$\Gamma_{ij;lm}^{||}(\vec{k},\vec{p}|\vec{q},\vec{s})$ and
$\Gamma_{ij;lm}^{\bot}(\vec{k},\vec{p}|\vec{q},\vec{s})$ are the
initial amplitudes. The quantity
\begin{eqnarray}
\Gamma_{ij;lm}^{||}&&(\vec{k},\vec{p}|\vec{q},\vec{s})\nonumber\\
&&= \frac12\Bigl(V_{ij;lm}(\vec{k},\vec{p}|\vec{q},\vec{s})
+V_{ji;ml}(\vec{p},\vec{k}|\vec{s},\vec{q})\nonumber\\
&&+G^{(1)}_{ij;lm}(\vec{k},\vec{p}|\vec{q},\vec{s})
+G^{(1)}_{ji;ml}(\vec{p},\vec{k}|\vec{s},\vec{q})\nonumber\\
&&+G^{(3)}_{ij;lm}(\vec{k},\vec{p}|\vec{q},\vec{s})
+G^{(3)}_{ji;ml}(\vec{p},\vec{k}|\vec{s},\vec{q})\nonumber\\
&&+G^{(4)}_{ij;lm}(\vec{k},\vec{p}|\vec{q},\vec{s})
+G^{(4)}_{ji;ml}(\vec{p},\vec{k}|\vec{s},\vec{q})\Bigr),\\
V_{ij;lm}&&(\vec{k},\vec{p}|\vec{q},\vec{s})=V_1\Bigl(
u_{\vec{q}-\vec{p}} w_{i1}(\vec{k}) w_{j3}(\vec{p})
w^*_{l3}(\vec{q}) w^*_{m1}(\vec{s})\nonumber\\
&&+u_{\vec{q}-\vec{p}}^* w_{i2}(\vec{k}) w_{j4}(\vec{p})
w^*_{l4}(\vec{q}) w^*_{m2}(\vec{s})\Bigr)\nonumber\\
&&+\frac{V_2}{2}\sum_{r=1}^4 f_{\vec{q}-\vec{p}}w_{ir}(\vec{k})
w_{jr}(\vec{p}) w^*_{lr}(\vec{q}) w^*_{mr}(\vec{s}),
\end{eqnarray}
\begin{eqnarray}
G^{(1)}_{ij;lm}&&(\vec{k},\vec{p}|\vec{q},\vec{s})=G_1
w_{i1}(\vec{k}) w_{j2}(\vec{p}) w^*_{l2}(\vec{q})
w^*_{m1}(\vec{s}),\\
G^{(3)}_{ij;lm}&&(\vec{k},\vec{p}|\vec{q},\vec{s})\nonumber\\
&&=G_3 u_{\vec{q}-\vec{p}} w_{i3}(\vec{k}) w_{j4}(\vec{p})
w^*_{l4}(\vec{q})w^*_{m3}(\vec{s}),\\
G^{(4)}_{ij;lm}&&(\vec{k},\vec{p}|\vec{q},\vec{s})=G_4\Bigl(
u_{\vec{q}-\vec{p}}^* w_{i1}(\vec{k}) w_{j4}(\vec{p})
w^*_{l4}(\vec{q}) w^*_{m1}(\vec{s})\nonumber\\
&&+u_{\vec{q}-\vec{p}} w_{i2}(\vec{k}) w_{j3}(\vec{p})
w^*_{l3}(\vec{q}) w^*_{m2}(\vec{s})\Bigr)
\end{eqnarray}
corresponds to the intensity of the interaction of fermions with
parallel spin projections, while the quantity
\begin{eqnarray}
\Gamma_{ij;lm}^{\bot}&&(\vec{k},\vec{p}|\vec{q},\vec{s})
=U_{ij;lm}(\vec{k},\vec{p}|\vec{q},\vec{s})
\nonumber\\
&&+V_{ij;lm}(\vec{k},\vec{p}|\vec{q},\vec{s})+
V_{ji;ml}(\vec{p},\vec{k}|\vec{s},\vec{q})\nonumber\\
&&+G^{(1)}_{ij;lm}(\vec{k},\vec{p}|\vec{q},\vec{s})
+G^{(1)}_{ji;ml}(\vec{p},\vec{k}|\vec{s},\vec{q})\nonumber\\
&&+G^{(3)}_{ij;lm}(\vec{k},\vec{p}|\vec{q},\vec{s})
+G^{(3)}_{ji;ml}(\vec{p},\vec{k}|\vec{s},\vec{q})\nonumber\\
&&+G^{(4)}_{ij;lm}(\vec{k},\vec{p}|\vec{q},\vec{s})
+G^{(4)}_{ji;ml}(\vec{p},\vec{k}|\vec{s},\vec{q}),\\
U_{ij;lm}&&(\vec{k},\vec{p}|\vec{q},\vec{s})
\nonumber\\
&&=U\sum_{r=1}^4 w_{ir}(\vec{k}) w_{jr}(\vec{p}) w^*_{lr}(\vec{q})
w^*_{mr}(\vec{s})
\end{eqnarray}
describes the interaction of fermions with antiparallel spin
projections. Indices ${i,j,l,m}$ correspond to the number of the
energy band and acquire the values 1, 2, 3, or 4.

\section{Effective interaction and equation for the superconducting order
parameter}

\begin{figure}
\begin{center}
\resizebox{0.41\textwidth}{!}{%
  \includegraphics{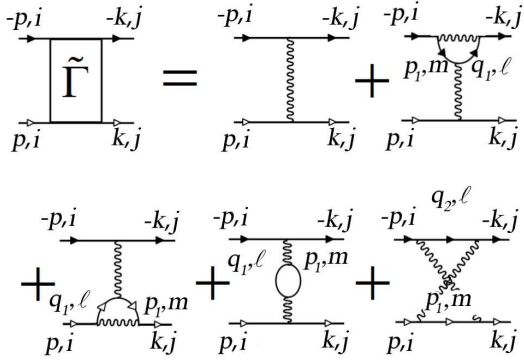}
} \caption{The sum of diagrams for the effective interaction of
electrons in the Cooper channel for graphene bilayer. Solid lines
correspond to Green's functions for electrons with spin
projections $+{\textstyle{1 \over2}}$ and $-{\textstyle{1 \over
2}}$ and energies corresponding to the energy bands $E_{i}$,
$E_{j}$, $E_{l}$, and $E_{m}$. Subscripts $i$ and $j$ acquire the
values of 1 or 2, whereas $l$ and $m$ acquire the values of 1, 2,
3, or 4. Momenta $\vec{q}_1$ and $\vec{q}_2$ are defined by
(\ref{q1q2}).}\label{diagrams}
\end{center}
\end{figure}

In this paper, we use the Born weak-coupling approximation, in
which the hierarchy of model parameters has the form
\begin{equation}\label{hierarchy}
W>U>V_1>V_2>G_1>G_3,\,G_4,
\end{equation}
where $W$ is the bandwidth in graphene bilayer (\ref{spectra}). In
the calculation of the scattering amplitude in the Cooper channel,
the condition (\ref{hierarchy}) allows us to limit the
consideration to only the second-order diagrams in the effective
interaction of two electrons with opposite values of the momentum
and spin and use the quantity $\widetilde{\Gamma}(\vec{p},
\vec{k})$ for it. Figure~\ref{diagrams} depicts the sum of
diagrams which determines $\widetilde{\Gamma}(\vec{p}, \vec{k})$.
Here, solid lines correspond to Green's functions for the
electrons with opposite spin projections $+\frac12$ (light arrows)
and $-\frac12$ (black arrows). The first diagram describes the
initial interaction of two electrons in the Cooper channel. Here,
the wavy lines correspond to the initial interaction. The next
four diagrams in Fig.~\ref{diagrams} correspond to the
second-order scattering processes
$\delta\widetilde{\Gamma}(\vec{p}, \vec{k})$ and describe the
polarization effects of the filled Fermi sphere. In the diagrams,
the presence of solid lines without arrows means the summation
over the spin projections values.

The possibility of the Cooper pairing is determined by the
features of the energy structure and the effective interaction of
electrons near the Fermi level~\cite{Gor'kov61}. If we assume that
the chemical potential in doped graphene bilayer is located in the
two upper bands $E_{1\vec{k}}$ and $E_{2\vec{k}}$
(Fig.~\ref{two_contours}a), we can consider the situation in which
the initial and final momenta of electrons in the Cooper channel
also belong to the two upper bands and analyze the conditions for
the Kohn-Luttinger superconducting pairing. At that, indices $i$
and $j$ in the diagrams (Fig.~\ref{diagrams}) will acquire the
values of 1 or 2.

Introducing the analytical expressions for the diagrams, we get
the effective interaction in the form
\begin{eqnarray}\label{Gamma_wave}
&&\widetilde{\Gamma}(\vec{p},\vec{k})=
\widetilde{\Gamma}_0(\vec{p},\vec{k})
+\delta\widetilde{\Gamma}(\vec{p},\vec{k}),\\
&&\widetilde{\Gamma}_0(\vec{p},\vec{k})=\Gamma^{\bot}_{ii;jj}(\vec{p},
-\vec{p}| -\vec{k},
\vec{k}),\\
&&\delta \tilde \Gamma (\vec{p},\vec{k})=
\frac{1}{N}\sum\limits_{l,m, \vec{p}_1}
\Gamma^{\bot}_{il;jm}(\vec{p}, \vec{q}_2| -\vec{k},
\vec{p}_1)\\
&&\times\Gamma^{\bot}_{mi;lj}(\vec{p}_1,
-\vec{p}|\vec{q}_2,\vec{k})
\chi_{l,m}(\vec{q}_2, \vec{p}_1)\nonumber\\
&&+\frac{2}{N}\sum\limits_{l,m, \vec{p}_1}  \Bigl\{
\Gamma^{\bot}_{im;lj}( \vec{p}, \vec{p}_1| \vec{q}_1,
\vec{k})\nonumber\\
&&\times\left[\Gamma^{||}_{li;mj}( \vec{q}_1, -\vec{p}| \vec{p}_1,
-\vec{k}) -
\Gamma^{||}_{li;jm}( \vec{q}_1, -\vec{p}| -\vec{k}, \vec{p}_1) \right] \Bigr.\nonumber\\
&&+\Bigl.\Gamma^{\bot}_{li;jm}(\vec{q}_1, -\vec{p}| -\vec{k}, \vec{p}_1)\nonumber\\
&&\times\left[\Gamma^{||}_{im;jl}( \vec{p}, \vec{p}_1| \vec{k},
\vec{q}_1) - \Gamma^{||}_{im;lj}( \vec{p}, \vec{p}_1| \vec{q}_1,
\vec{k})\right] \Bigr\}\chi_{l,m}(\vec{q}_1,\vec{p}_1).\nonumber
\end{eqnarray}
Here, we use the notations for the generalized susceptibilities
\begin{equation}
\chi_{l,m}(\vec{k},\vec{p}) = \frac{f(E_{l\vec{k}}) -
f(E_{m\vec{p}})} {E_{m\vec{p}} - E_{l\vec{k}}},
\end{equation}
where $f(x)=(\exp(\frac{x-\mu}{T})+1)^{-1}$ is the Fermi-Dirac
function and the energies $E_{i\vec{k}}$ are defined by the
expressions~(\ref{spectra}). Additionally, we have introduced the
following notations for the combinations of the momenta
\begin{equation}\label{q1q2}
\vec{q}_1 =  \vec{p}_1 + \vec{p} - \vec{k},\qquad \vec{q}_2 =
\vec{p}_1-\vec{p}-\vec{k}.
\end{equation}

The renormalized expression for the effective interaction allows
us to analyze the conditions for the occurrence of
superconductivity in the system. It is known~\cite{Gor'kov61} that
the development of the Cooper instability can be established from
the consideration of the homogeneous part of the Bethe-Salpeter
equation. At that, the dependence of the scattering amplitude
$\Gamma(\vec{p},\vec{k})$ on momentum $\vec{k}$ is factorized and
we get the integral equation for the superconducting order
parameter $\Delta(\vec{p})$. After the integration over the
isoenergetic contours, the problem of the Cooper instability can
be reduced to the eigenvalue
problem~\cite{Scalapino86,Baranov92,Hlubina99,Raghu10,Alexandrov11}
\begin{equation}
\label{IntegralEqPhi}
\frac{3\sqrt{3}}{8\pi^2}\oint\limits_{\varepsilon_{\vec{q}}=\mu}
\frac{d\hat{\vec{q}}} {v_F(\hat{\vec{q}})}
\widetilde{\Gamma}(\hat{\vec{\vec{p}}},\hat{\vec{q}})
\Delta(\hat{\vec{q}})=\lambda\Delta(\hat{\vec{p}}),
\end{equation}
where the eigenvector is the superconducting order parameter
$\Delta(\hat{\vec{q}})$ and the eigenvalues $\lambda$ satisfy the
relation $\lambda^{-1}\simeq \ln(T_c/W)$. Here, the momenta
$\hat{\vec{p}}$ and $\hat{\vec{q}}$ belong to the Fermi surface
and $v_F(\hat{\vec{q}})$ is the Fermi velocity. Equation
(\ref{IntegralEqPhi}) is solved in accordance with the common
scheme described in~\cite{Kagan14,Kagan14a}. The integration is
fulfilled with the allowance for the fact that the Fermi contour
near each Dirac point consists of two lines
(Fig.~\ref{two_contours}b).

\section{Results and discussion}

Let us consider the phase diagram of the superconducting state of
the graphene bilayer and the modifications of this diagram in the
different regimes obtained by solving Eq. ($\ref{IntegralEqPhi}$).
When building the phase diagram, we divided the multisheet Fermi
contour into 180 intervals and the Brillouin zone of the graphene
bilayer, into $5\cdot10^4$ cells. It was established that the
chosen method of division is sufficient for the correct
description of the dependence of the effective coupling constant
$\lambda$ on the electron density $n$~\cite{Kagan14}. Based on the
obtained dependences $\lambda(n)$ for different values of the
intersite $V_1$ and interplane $G_1,\,G_3$ and $G_4$ Coulomb
interactions, we built the phase diagrams of the Shubin-Vonsovsky
model for bilayer graphene, which reflect the competition between
the superconducting phases with different types of symmetry of the
order parameter.

So far, there has been no agreement regarding the values of
parameters of the intra- and interplanar Coulomb interactions in
the graphene bilayer. The ab initio calculations for
graphite~\cite{Wehling11} showed that the value of Hubbard
repulsion is $U=8.0\,\textrm{eV}$, which is consistent with the
estimation made in~\cite{Levin74} and contradicts the intuitively
expected small value of $U$ and weak-coupling limit $U<W$ (it is
known~\cite{Reich02} that $t_1\approx2.8\,\textrm{eV}$). The
authors of~\cite{Wehling11} calculated the parameters of Coulomb
repulsion between electrons of the nearest and the next-to-nearest
carbon atoms: $V_1=3.9\,\textrm{eV}$ and $V_2=2.4\,\textrm{eV}$,
respectively. At the same time, the other authors (see, for
example,~\cite{Perfetto07}) consider these parameters to be much
smaller. The authors of~\cite{Milovanovic12} mentioned that the
estimation of the parameters of Coulomb interaction, including the
Hubbard repulsion, in the graphene bilayer strongly depends on the
calculation scheme which is used. In our calculation, we apply the
parameter hierarchy (\ref{hierarchy}), which allows us to use the
Born weak-coupling approximation. For interlayer hopping
parameters $\gamma_1$ and $\gamma_3$, we use the values similar to
those determined in~\cite{Dresselhaus02,Brandt88} for graphite.
\begin{figure}
\begin{center}
\resizebox{0.48\textwidth}{!}{%
  \includegraphics{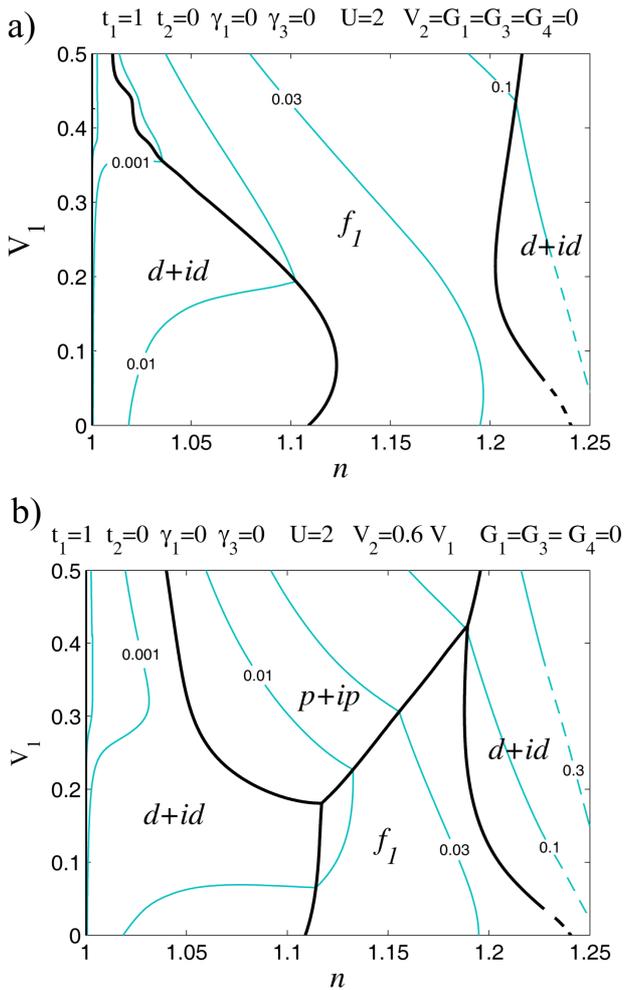}
} \caption{Phase diagram of the superconducting state of the
graphene bilayer shown as a function of the variables "$n-V_1$" at
$t_2=0,\,\gamma_1=\gamma_3=0,\,U=2,\,G_1=G_3=G_4=0$ for (a)
$V_2=0$ and (b) $V_2=0.6V_1$ (all the parameters are in units of
$|t_1|$). For all the points on the same thin blue line, the value
of $|\lambda|$ is constant and marked with the corresponding
number.}\label{PD_limiting}
\end{center}
\end{figure}

\begin{figure}
\begin{center}
\resizebox{0.48\textwidth}{!}{%
  \includegraphics{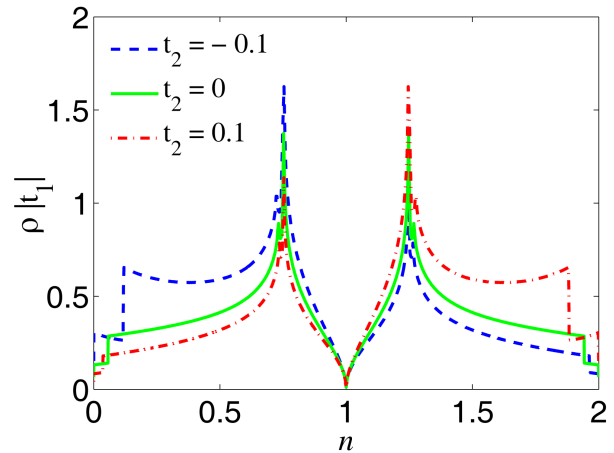}
} \caption{Modification of the dependence of the density of
electron states of the graphene bilayer from electron density
calculated for atoms in the same layer with respect to electron
hoppings to the next-to-nearest neighboring
atoms.}\label{DOS_bilayer}
\end{center}
\end{figure}

First, let us consider the limiting case when the bilayer energy
spectrum is described by the only one hopping parameter
($t_1\neq0,\,t_2=\gamma_1=\gamma_3=0$). The Hubbard repulsion is
also taken into account $U=2$ (hereinafter, all the parameters are
given in units of $|t_1|$). The Coulomb repulsion between
electrons ($V_1\neq0$) of the neighboring carbon atoms in the same
layer is taken into account as well. At the same time, the
interlayer Coulomb interactions are not taken into account
($G_1=G_3=G_4=0$). Thus, in the chosen regime, the graphene
bilayer consists of two isolated single layers. The phase diagram
of the superconducting state shown as a function of the variables
"$n-V_1$" for this case is presented in Fig.~\ref{PD_limiting}a.
It can be seen that the phase diagram comprises three regions. At
low electron densities $n$, the ground state of the system
corresponds to the superconductivity with the $d+id-$wave symmetry
of the order parameter, which is described by the 2D
representation $E_2$, the contribution to which is determined by
the harmonics
\begin{eqnarray}\label{E2}
g_{m}^{(d+id)}(\phi)=\frac{1}{\sqrt{\pi}}\,(A\,\textrm{sin}\,
(2m+2)\phi+B\,\textrm{cos}\,(2m+2)\phi)\nonumber,
\end{eqnarray}
where subscripts $m$ run over the values for which the
coefficients $(2m+2)$ are not multiples of 3. At the intermediate
electron densities, the superconducting $f-$wave pairing is
implemented, the contribution to which is determined by the
harmonics $g_{m}^{(f_1)}(\phi)=\displaystyle\frac{1}{\sqrt{\pi}}\,
\textrm{sin}\,(6m+3)\phi$ (here $m\in[\,0,\infty)$), while the
contribution of the harmonics
$g_{m}^{(f_2)}(\phi)=\displaystyle\frac{1}{\sqrt{\pi}}\,
\textrm{cos}\,(6m+3)\phi$ is absent. At the large values of $n$,
the domain of the superconducting $d+id-$wave pairing
occurs~\cite{Nandkishore12}. With the increase of the parameter
$V_1$ of the intersite Coulomb interaction, in the region of small
values of $n$, the $d+id-$wave pairing is suppressed and the
pairing with the $f-$wave symmetry of the order parameter is
implemented. Thin blue lines in Fig.~\ref{PD_limiting} are the
lines of the equal values of the effective coupling constant
$|\lambda|$. It can be seen that in this case in the proximity of
the Van Hove filling $n_{VH}$ (solid curve in
Fig.~\ref{DOS_bilayer}) the effective coupling constant attains
the values $|\lambda|=0.1$.

It should be noted that to avoid the summation of the parquet
diagrams~\cite{Dzyaloshinskii88a,Dzyaloshinskii88b,Zheleznyak97},
we do not analyze here the electron density regions that are very
close to the Van Hove singularity in the density of electron
states of bilayer graphene (Fig.~\ref{DOS_bilayer}). For this
reason, the boundaries between different domains of the
implementation of the Kohn-Luttinger superconducting pairing, as
well as the lines of the equal value of $|\lambda|$ that are very
close to the Van Hove singularity are indicated in the phase
diagram by the dashed lines.
\begin{figure}
\begin{center}
\resizebox{0.48\textwidth}{!}{%
  \includegraphics{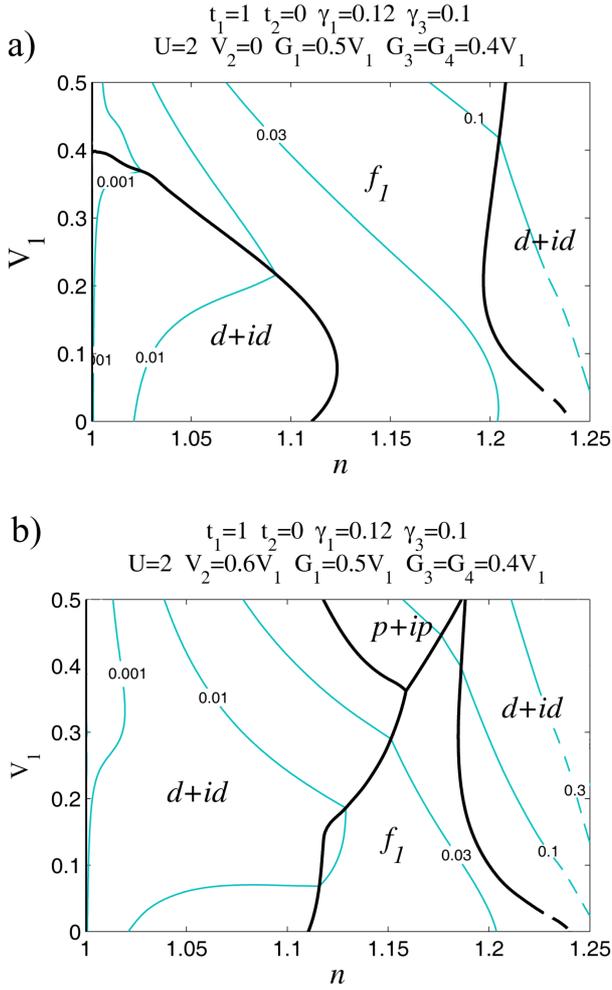}
} \caption{Phase diagram of the superconducting state of the
graphene bilayer shown as a function of the variables "$n-V_1$"
for
$t_2=0,\,\gamma_1=0.12,\,\gamma_3=0.1,\,U=2,\,G_1=0.5V_1,\,G_3=G_4=0.4V_1$,
at (a) $V_2=0$ and (b) $V_2=0.6V_1$ (all the parameters are given
in units of $|t_1|$). Thin blue curves are the lines of the
constant values of $|\lambda|$.}\label{PD_G3G4}
\end{center}
\end{figure}

Thus, in the numerical calculation for the graphene bilayer for
the chosen parameters, we made the limiting transition to the
results obtained by us previously for the graphene
monolayer~\cite{Kagan14,Kagan14a}.

Let us consider the modification of the phase diagram for the
isolated graphene single layers with regard to the long-range
intraplane Coulomb interactions between electrons $V_2$. It can be
seen in Fig.~\ref{PD_limiting}b for the fixed ratio between the
parameters of the long-range Coulomb interactions $V_2=0.6V_1$
that when  $V_2$ is taken into account, the phase diagram changes
qualitatively. This change involves the suppression of a large
domain of the superconducting state with the $f-$wave symmetry at
the intermediate electron densities and the implementation of the
superconducting pairing with the $p+ip-$wave symmetry of the order
parameter. In addition, when $V_2$ is taken into account, the
effective coupling constant increases to the value
$|\lambda|=0.3$.

Now, let us consider the modification of the phase diagram of the
superconducting state with respect to the interplanar
interactions. When the interlayer electron hoppings
$\gamma_1=0.12$ and $\gamma_3=0.1$ are taken into account while
the other parameters being the same as in Fig.~\ref{PD_limiting},
the phase diagram of the graphene bilayer remains nearly
unchanged.

Inclusion of the Coulomb interaction $G_1$ in the consideration
weakly shifts the boundaries of the $f_1-$wave and $d+id-$wave
pairing in the phase diagram in Fig.~\ref{PD_limiting} and does
not affect the absolute values of $\lambda$. Figure~\ref{PD_G3G4}
shows the effect of taking into account the interlayer Coulomb
interactions $G_3$ and $G_4$. Figure~\ref{PD_G3G4}a shows the
phase diagram of the Shubin-Vonsovsky model for the graphene
bilayer for the set of parameters
$t_2=0,\,\gamma_1=0.12,\,\gamma_3=0.1,\,U=2$ and $V_2=0$ for the
chosen ratios between the interlayer and intersite Coulomb
interactions $G_1=0.5V_1,\,G_3=G_4=0.4V_1$, according to the
hierarchy of the parameters (\ref{hierarchy}). The calculation
shows that the separate increase of the parameters $G_3$ and $G_4$
suppresses the $d+id-$wave pairing and, at the same time, broadens
the $f-$wave pairing region at small electron densities. The
superconducting $d+id-$phase is suppressed the most effectively by
enhancing the parameter $G_4$ of the interlayer Coulomb
interaction. When the interactions $G_3$ and $G_4$ are
simultaneously taken into account (Fig.~\ref{PD_G3G4}a), then
along with the intensive suppression of the superconducting
$d+id-$wave pairing at small electron densities and the
implementation of the superconductivity with the $f-$wave symmetry
of the order parameter, the growth of the absolute values of
effective coupling constant $\lambda$ is also observed.

Figure~\ref{PD_G3G4}b depicts the phase diagram of the graphene
bilayer calculated for the same parameters as in
Fig.~\ref{PD_G3G4}a but with respect to the long-range intraplane
Coulomb repulsion between electrons $V_2$. Comparison of
Figs.~\ref{PD_G3G4}b and ~\ref{PD_limiting}b shows that the
account for $G_3\neq0$ and $G_4\neq0$ leads to the strong
competition between the $d+id-$wave and $p+ip-$wave pairings with
the significant suppression of the $p+ip-$wave pairing in the
region of the intermediate electron densities. In this case, in
the remained region of the $p+ip-$wave pairing, $|\lambda_{p+ip}|$
slightly exceeds $|\lambda_f|$.

\begin{figure}
\begin{center}
\resizebox{0.48\textwidth}{!}{%
  \includegraphics{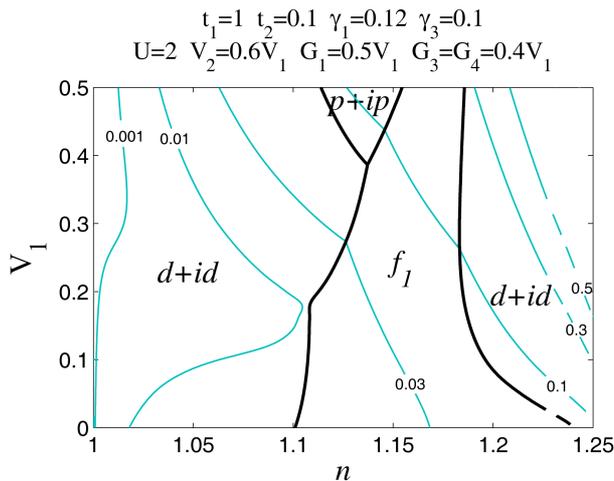}
} \caption{Phase diagram of the superconducting state of the
graphene bilayer shown as a function of the variables "$n-V_1$" at
$t_2=0.1,\,\gamma_1=0.12,\,\gamma_3=0.1,\,U=2,\,G_1=0.5V_1,\,G_3=G_4=0.4V_1$
(all the parameters are in units of $|t_1|$). Thin blue curves are
the lines of the constant values of $|\lambda|$.}\label{PD_t2}
\end{center}
\end{figure}

The account for electron hoppings to the next-to-nearest carbon
atoms $t_2$ does not qualitatively affects the competition between
the superconducting phases (Fig.~\ref{PD_G3G4}).
Figure~\ref{PD_t2} depicts the phase diagram of the graphene
bilayer obtained for the parameters
$t_2=0.1,\,\gamma_1=0.12,\,\gamma_3=0.1,\,U=2,\,G_1=0.5V_1$ and
$G_3=G_4=0.4V_1$. Such a behavior of the system is explained by
the fact that switching on of the hoppings $t_2>0$ or $t_2<0$ for
the graphene bilayer, similarly to the case of the monolayer
investigated by us in~\cite{Kagan14,Kagan14a}, does not
significantly modify the density of electron states in the carrier
concentration regions between the Dirac point and both points
$n_{VH}$ (Fig.~\ref{DOS_bilayer}). However, it can be seen in
Fig.~\ref{PD_t2} that the account for the hoppings $t_2$ leads to
an increase of the effective interaction in the absolute values
and, consequently, to the higher superconducting transition
temperatures in an idealized graphene bilayer.

It should be noted that the Kohn--Luttinger superconductivity in
the graphene single layer and bilayer never develops near the
Dirac points. The calculations show that in the vicinity of these
points, where the linear approximation for the energy spectrum of
the graphene single layer and the parabolic approximation for the
spectrum of the graphene bilayer work pretty well, the density of
states is very low and the effective coupling constant
$|\lambda|<10^{-2}$. The higher values of $|\lambda|$, which are
indicative of the development of the Cooper instability, arise at
the electron densities $n>1.15$. However, at such densities, the
energy spectrum of the bilayer along the direction $KM$ of the
Brillouin zone (Fig.~\ref{two_contours}b) already significantly
differs from the Dirac approximation.
%\begin{eqnarray}
%&&\gamma_1=0.12,\,\gamma_3=0.1,\,U=2,\,V_1=0\div0.5,\nonumber\\
%&&G_1=0.5V_1,\,G_3=G_4=0.4V_1
%\end{eqnarray}

\section{Conclusions}

In the work, we have analyzed the conditions for the
Kohn-Luttinger superconductivity in a semimetal with the Dirac
spectrum using as an example an idealized graphene bilayer,
disregarding the van der Waals potential of the substrate and both
magnetic and non-magnetic impurities. The electronic structure of
graphene bilayer is described in the Shubin-Vonsovsky model taking
into account not only the Coulomb repulsion of electrons of the
same carbon atom, but also the intersite and interlayer Coulomb
interactions. It was shown that in such a system, the
Kohn-Luttinger polarization contributions lead to the effective
attraction between electrons in the Cooper channel. The
constructed superconducting phase diagram of the system determines
the Cooper pairing domains with the different types of the
symmetry of the order parameter, depending on the intersite
Coulomb interactions and the electron densities. The analysis of
the phase diagram showed that the inclusion of the Kohn-Luttinger
renormalizations up to the second order of perturbation theory
inclusively and the allowance for the long-range Coulomb
interactions $V_1$ and $V_2$ determine, to a considerable extent,
the competition between the superconducting phases with the
$f-$wave, $p+ip-$wave, and $d+id$-wave types of the symmetry of
the order parameter. They also lead to a significant increase in
the absolute values of the effective interaction. It was shown
that the allowance for the interlayer Coulomb interactions $G_3$
and $G_4$, as well as for the distant electron hoppings $t_2$,
leads to an additional increase in the effective interaction and,
hence, to the higher superconducting transition temperatures in an
idealized graphene bilayer.

Our calculation showed that the Kohn-Luttinger mechanism can lead
to the superconducting transition temperatures $T_c\sim
20\div40~K$ in an idealized graphene bilayer. Contrary to these
rather optimistic estimations, in real graphene, as it was
mentioned in Introduction, superconductivity has not been found
yet. This material is only close to superconductivity.

For a few reasons, the results of the theoretical calculations
reported here can differ from the experimental situation. First,
we did not take into account the effect of the van der Waals
potential of the
substrate~\cite{Gomez09,Bostrom12,Klimchitskaya13}. It seems that
the effect of this potential should be weakened with the increase
of number of layers. However, even in the multilayer systems the
van der Waals forces can degrade the conditions for the
development of the Cooper instability.

Second, as we mentioned in Section 4, there has been no agreement
regarding the values of the parameters of the intraplane and
interplanar Coulomb interactions in the graphene bilayer in the
literature. In this work, we used the values of the intraplane
Coulomb interactions that are close to those obtained from the ab
initio calculation in~\cite{Wehling11} for graphite. The values of
the interplanar Coulomb interactions were chosen to satisfy the
hierarchy of the parameters of the Born weak-coupling
approximation.

Third, in our calculations, we considered a pure graphene bilayer
with the ideal structure, whereas the real material contains
numerous impurities and structural defects. It is well known that,
in contrast to the traditional $s-$wave pairing, for the anomalous
pairing with the $f$-wave, $p+ip$-wave, and $d+id$-wave symmetries
of the order parameter, nonmagnetic impurities and structural
defects can destroy the superconducting order~\cite{Black14b}.

In addition, we should mention one more possible reason for the
discrepancy between the theoretical calculations on
superconductivity in graphene and the experimentally observed
situation. In recent paper~\cite{Kats14}, the effect of quantum
fluctuations ($T=0$) on the graphene layers was investigated. It
was shown that these fluctuations initiate the logarithmic
corrections to the moduli of elasticity and bending of the layers.
In other words, according to~\cite{Kats14}, the quantum
fluctuations connected with the bending vibrations of the graphene
layers can lead to the situation when the electrons do not move
along the atomically smooth layers but along the strongly curved
string-like trajectories, as in quantum chromodynamics. This
situation requires further investigations, although in this case
the superconductivity is not at all excluded and even can be
enhanced by the exchange of bending vibration quanta between the
pairing electrons.

We thank V.V. Val'kov for useful discussions. This work is
supported by the Russian Foundation for Basic Research (projects
nos. 14-02-00058 and 14-02-31237). One of the authors
(M.\,Yu.\,K.) gratefully acknowledges support from the Basic
Research Program of the National Research University Higher School
of Economics. Another one (M.\,M.\,K.) thanks the scholarship
SP-1361.2015.1 of the President of the Russian Federation and the
Dynasty foundation.

\end{document}